\documentclass[journal]{IEEEtran}
\usepackage[table,xcdraw]{xcolor}
\usepackage{amsmath,amssymb,amsfonts}
\usepackage{algorithmic}
\usepackage{graphicx}
\usepackage{textcomp}
\usepackage{booktabs,tabu}
\usepackage{soul}

\ifCLASSINFOpdf
\else
\fi
\begin{document}
%
\title{Effect of Kinematics and Fluency in Adversarial Synthetic Data Generation for ASL Recognition with RF Sensors}
%
%
%

\author{M.M. Rahman, \IEEEmembership{Student Member, IEEE}, E. Malaia, A.C. Gurbuz, \IEEEmembership{Senior Member, IEEE},  D.J. Griffin, C. Crawford and S.Z. Gurbuz, \IEEEmembership{Senior Member, IEEE}
\thanks{This work was supported in part by the National Science Foundation under Grants 1932547, 1931861 and 1734938. Human studies research was conducted under University of Alabama Institutional Review Board (IRB) Protocol \#18-06-1271. }
\thanks{M.M. Rahman and S.Z. Gurbuz are with the University of Alabama, Department of Electrical and Computer Engineering, Tuscaloosa, AL 35487 (email: mrahman17@crimson.ua.edu, szgurbuz@ua.edu).}
\thanks{E. Malaia is with the University of Alabama, Department of Communication Disorders, Tuscaloosa, AL 35487 (e-mail: eamalaia@ua.edu).}
\thanks{A.C. Gurbuz is with Mississippi State University, Department of Electrical and Computer Engineering (e-mail: gurbuz@ece.msstate.edu).}
\thanks{D.J. Griffin is with the University of Alabama, Department of Communication Studies, Tuscaloosa, AL 35487 (e-mail: djgriffin1@ua.edu).}
\thanks{C. Crawford is with the University of Alabama, Department of Computer Science, Tuscaloosa, AL 35487 (e-mail: crawford@ua.edu).}
}

%
%

\markboth{Journal of \LaTeX\ Class Files,~Vol.~14, No.~8, August~2015}%
{Shell \MakeLowercase{\textit{et al.}}: Bare Demo of IEEEtran.cls for IEEE Journals}
%



\maketitle

\begin{abstract}
RF sensors have been recently proposed as a new modality for sign language processing technology.  They are non-contact, effective in the dark, and acquire a direct measurement of signing kinematic via exploitation of the micro-Doppler effect.  First, this work provides an in depth, comparative examination of the kinematic properties of signing as measured by RF sensors for both fluent ASL users and hearing imitation signers.  Second, as ASL recognition techniques utilizing deep learning requires a large amount of training data, this work examines the effect of signing kinematics and subject fluency on adversarial learning techniques for data synthesis. Two different approaches for the synthetic training data generation are proposed:  1) adversarial domain adaptation to minimize the differences between imitation signing and fluent signing data, and 2) kinematically-constrained generative adversarial networks for accurate synthesis of RF signing signatures.  The results show that the kinematic discrepancies between imitation signing and fluent signing are so significant that training on data directly synthesized from fluent RF signers offers greater performance (93\% top-5 accuracy) than that produced by adaptation of imitation signing (88\% top-5 accuracy) when classifying 100 ASL signs.  
\end{abstract}

\begin{IEEEkeywords}
radar, micro-Doppler, sign language, ASL, adversarial learning, kinematics
\end{IEEEkeywords}

%
\IEEEpeerreviewmaketitle

\section{Introduction}
\label{sec:introduction}
According to the World Federation for the Deaf (WFD), an estimated 74 million people world-wide communicate using sign language.  American Sign Language (ASL) is estimated to be the primary mode of communication for over a million people in North America and Canada, based on statistics provided by Gallaudet University (the world's only university designed to be barrier-free for deaf and hard of hearing students located in Washington, D.C.).  Although much research in ASL recognition has focused on translation (e.g. sign to speech) as a means to bridge the communication gap between Deaf and hearing individuals, this work aims at the development of RF sensing-enabled sign language processing (SLP) technologies in an ambient, non-invasive fashion for human-computer interaction (HCI) applications, such as assistive robots \cite{Robots1,Robots2} and smart environments \cite{Robots3}.  RF sensors have been successfully used for remote health monitoring of vital signs \cite{ChangzhiLi2017}, fall detection \cite{Amin_SPM_Fall_2016}, gait analysis \cite{Seifert_TBME_2019}, and detection of  sleep apnea \cite{sleep} or sudden infant death syndrome \cite{baby}. The addition of ASL recognition capability to such systems would  extend their use potential to Deaf populations, and enhance the quality of life for those who use ASL.  

RF sensors offer unique advantages in that they are non-contact, not restrictive or invasive, operate at a distance, protect the privacy of the user and personal spaces, and are effective in the dark, regardless of what the individual is wearing.  Thus, RF sensors can recognize signing \cite{ASL_Patent2018,9187644} in situations where other sensors, such as wearables \cite{Wearable1,Wearable2} or cameras \cite{Video1,Kinect1,LeapMotion1}, are either undesireable or ineffective. RF sensors cannot perceive hand shapes or facial expressions, but they can provide a direct measure of distance and velocity as a function of time. Velocity can be obtained via the Doppler effect; namely, the principle that the frequency shift incurred in the received RF signal is proportional to the radial velocity of an object in motion.  While translational motion results in a central Doppler frequency,  \textit{micro-Doppler} \cite{ChenMD} refers to the frequency modulations generated about the center frequency that result from vibrations or rotations.  As such, the micro-Doppler signature is comprised of unique patterns directly related to the kinematics of the underlying motion, and can serve as a biometric to identify individuals \cite{Vandersmissen2018}, various activities \cite{Gurbuz_IET2017}, and gestures \cite{Arbabian2013,Wang2021,ASLR_JSens2020}. 

Deep learning has enabled great advances in the recognition capabilities for many sensing modalities, including RF sensors \cite{Gurbuz_SPM_2019}.  However, deep neural networks (DNNs) require large amounts of data to learn complex underlying data representations. In biomedical applications for human motion recognition,  acquisition of adequate sample sizes can be challenging;  targeted populations may be mobility-impaired or reluctant to participate.  Given that the Deaf are a minority community, its members are highly sought after for involvement in a variety of research studies, and thus may be wary of frequent requests. Inclusion of Deaf researchers is critical for hearing researchers to understand Deaf cultural perspectives and incorporate Deaf experiences and knowledge of the language into all aspects of SLP technology design \cite{ACM2019_ASL_Survey}.  Our research team includes a Child-of-Deaf-Adults (CODA), who is fluent in ASL, and benefits from the involvement of community partnerships with Gallaudet University and the Alabama Institute for the Deaf and Blind (AIDB), who have provided feedback on Deaf-centric design and aided in the recruitment of Deaf participants.

Nevertheless, reliance on extensive amounts of human subject data for training deep models can result in an undue burden on Deaf participants, even if well-intentioned and for the benefit of the community.  One way some researchers have addressed the need for signing data has been to utilize ASL learners or imitation signers, e.g. \cite{DeepASL2017,SignFi2018}, who are more easily recruitable.  Imitation signing refers to the process of asking hearing sign-naive participants to replicate the signs shown in any video. However, it can take learners of sign language at least 3 years to produce signs in a manner that is perceived as fluent by fluent signers \cite{beal2018hearing}.  Thus, even with ''training'' sessions to teach participants how to articulate the signs prior to conducting the experiment with imitation signing videos, the production of hearing imitation signers is not comparable to that of fluent signing and may indeed contain significant errors in temporal dynamics and repetitions (which RF sensors easily perceive), as well as hand shape and place of articulation, i.e. position of the sign in space. In our prior work \cite{ASLR_JSens2020,9114818}, we found that imitation signing and fluent signing occupy different regions in the feature space, enabling machine learning to 
effectively distinguish between imitation signing and fluent signing.

But does this mean that imitation signing data has no value in the training of DNNs for fluent ASL recognition?  Adversarial domain adaptation is an approach that has been utilized in the computer vision community for style transfer \cite{Style1,Style2,Style3} and image-to-image translation \cite{Liu2017UnsupervisedIT,Isola2017ImagetoImageTW,Zhu2017UnpairedIT}.  One approach to the design of generative adversarial networks (GANs) for domain adaptation is to use  Pix2Pix GAN \cite{Isola2017ImagetoImageTW}  for image-to-image translation  based on the conditional GAN, where a target image is generated, conditional on a given input image. In this case, the Pix2Pix GAN changes the loss function so that the generated image is both plausible in the content of the target domain, and is a plausible translation of the input image.  Another approach is to minimize the \textit{cycle-consistency loss} \cite{Zhu2017UnpairedIT,Kim2017LearningTD,Yi2017DualGANUD}, which enforces two mappings to be the reverse of each other: $F(G(x))=x$.  As an alternative to the cycle-consistency proposed with CycleGAN \cite{Zhu2017UnpairedIT}, TravelGAN \cite{Amodio2019TraVeLGANIT} has also been proposed, which instead utilizes an additional Siamese network to guide the generator in generating shared semantics, and thus learn mappings between more complex domains that have large differences beyond that of just style or texture.  Thus, one approach to training deep models for ASL recognition could be to utilize adversarial domain adaptation to transform the imitation signing signatures to have greater resemblance to fluent signing signatures.

A significant challenge to the application of adversarial learning to RF datasets, however, is that the data supplied to the DNNs are not optical images, but computed images, derived from time-frequency transform of the raw complex received RF signal.  Thus, the pixels in an RF micro-Doppler signature bear no relation to geometry, lighting, or perspective. In contrast, it is the kinematics of the human skeleton that determines the frequency profiles revealed in the RF signature.  In prior work, we have shown that consideration of kinematics can lead to great gains in DNN training for human activity recognition: 
\begin{enumerate}
    \item Data augmentation via temporal and spatial scaling of the underlying skeletal animation \cite{Seyfioglu_TAES_2019} yields much more effective, diversified and statistically independent samples than image-based data augmentation techniques, which can corrupt the signatures and result in physically impossible samples.
    
    \item  The classification accuracy for eight different daily human activities was boosted by 10\% simply by discarding 9,000 kinematically impossible samples, which were identified as outliers relative to real data samples using Principal Component Analysis (PCA) \cite{Erol_TAES_2020}, from a synthetic dataset of 40,000 samples generated using an Auxilliary Conditional GAN (ACGAN).
    \item Use of the signature envelope in addition to the signature itself in a multi-branch GAN (MBGAN) architecture was shown \cite{Erol_MBGAN_2020} to shift the distribution of synthetic human activity data so as to have greater overlap with that of real data, as visualized using t-SNE \cite{tsne}.  Studies of gross motor motion recognition showed that MBGAN offered improved classification accuracy \cite{Rahman_LRMBGAN_2021}.
\end{enumerate}
Thus, an alternative approach to the transformation of imitation signing signatures could be to directly synthesize ASL signatures for training using adversarial learning.  

In this work, we investigate the kinematic properties of sign production by fluent signers versus hearing imitation signers using RF sensors, as well as the impact of fluency and sign kinematics (i.e. components of sign phonology \cite{malaia2012kinematic, malaia2020syllable}) on training DNNs for classification of fluent signing using synthetic signatures that are (a) transformed from imitation signing data, versus (b) directly generated from a small set of real signatures from fluent signers.  In Section II, the experimental procedure for acquiring the 100-sign ASL datasets for imitation signers and fluent ASL users is presented.  The kinematic and linguistic properties of these signs, as listed in the ASL-LEX2 database \cite{ASL_LEX_2016}, are described.  In Section III, techniques for estimating these properties from the RF micro-Doppler signature for each sign are presented. In Section IV, the adversarial networks for domain adaptation and design of a 3-branch MBGAN for ASL training data synthesis are detailed.  In particular, using the estimators developed in Section III, the degree to which different adversarial networks preserve the salient properties of each sign are quantitatively evaluated and the advantages of embedding kinematics into the GAN architecture are demonstrated.  The similarity of transformed imitation signing and synthesized ASL is compared with that of fluent ASL signatures for different database sizes.  In Section IV, kinematically deviant signatures are sifted out from the generated data, and the resulting synthetic datasets are used to train DNNs.  In this way, the efficacy of the proposed approaches to classyfing a large number of ASL signs while minimizing real human subject data requirements is demonstrated.  The paper concludes in Section V with a discussion of results and future research directions.


\begin{table*}[t!]
\centering
\caption{Listing of the 100 ASL signs utilized in experiments.}
\includegraphics[height=8cm]{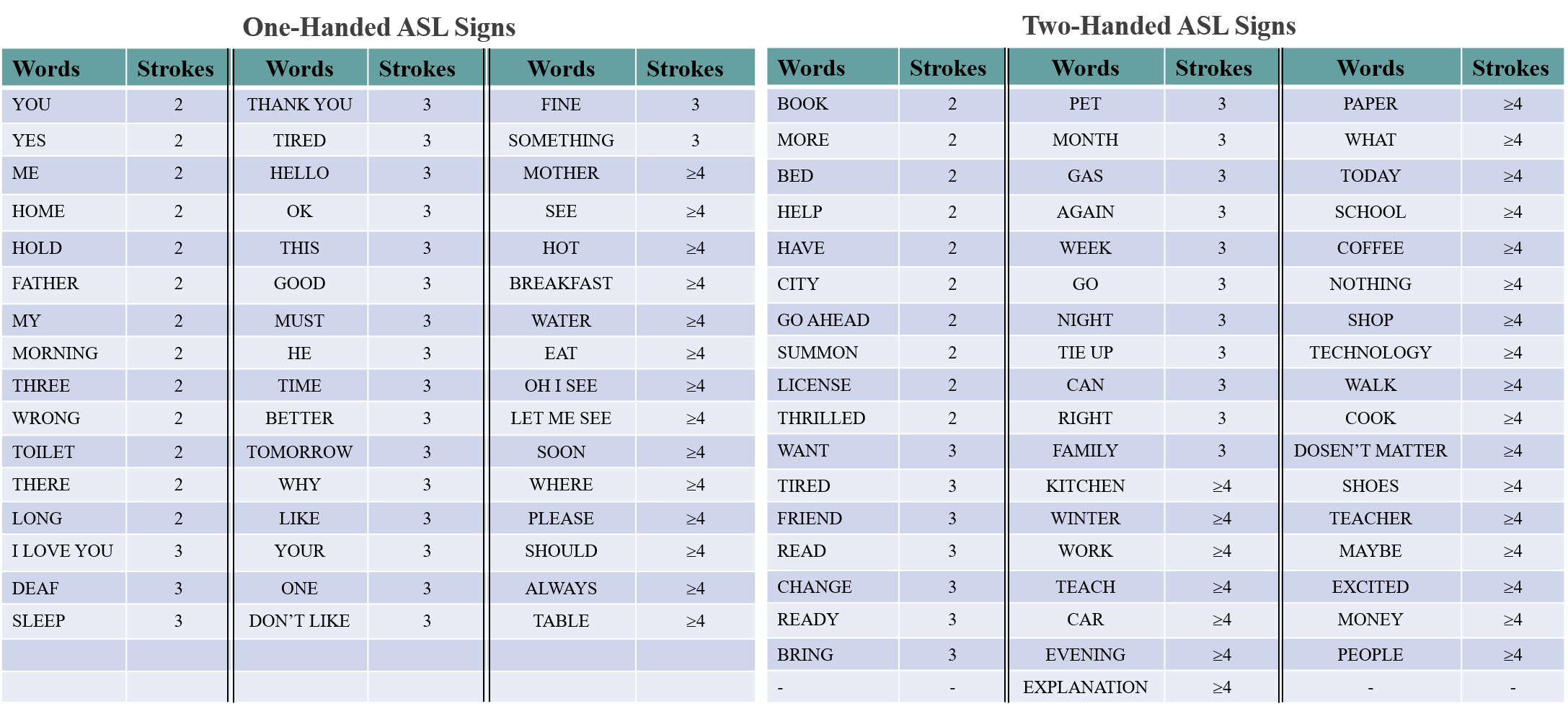} 
\label{fig: 100 ASL word}
\end{table*}

\section{Experimental RF Datasets}

\subsection{RF Sensors and Test Environment}
The RF sensor used in this work is a TI IWR1443BOOST 77 GHz - 81 GHz automotive short-range radar (SRR) sensor, which transmits a pulsed, linear frequency modulated continuous wave (FMCW) signal (a.k.a "chirp" signal). The normalized transmitted signal of the FMCW
radar is \cite{FMCW}
\begin{equation}
\label{eq:tx}
x_{tr}(t)=exp\bigg\{j2\pi(f_{c}t+\frac{k}{2}t^2)\bigg\}
\end{equation}
where $t$ denotes the fast time within a chirp (a frequency
modulation period), $T_{s}/2 \leq t \leq T_{s}/2$, $f_{c}$ and $k = B/T_{s}$ denote the center frequency and the frequency slope of the
chirp, and $B$ and $T_{s}$ denote the bandwidth and the time
duration of the chirp, respectively. 

The transmitted signal illuminates an ASL signer who sits 1.5 meters in front of the sensor and signs in ASL.  The radar receives backscatter from the moving arms and hands, as well as reflection from static parts of the body and environment.  According to geometric diffraction theory \cite{Keller62}, when
the wavelength of the incident wave is much smaller than the target size, the backscattered returns from the target can
be expressed as the superposition of a set of independent scattering centers. Thus, the  signal  received  by  the  receiver is a weighted summation of time-delayed, frequency-shifted versions of the transmitted signal given by the the superposition of returns from $M$ points on the body \cite{VanDorp2003}. Thus, 
\begin{equation}
\label{eq:rre}
x_{rec}(t) = \sum_{i=1}^{M}{a_{i}exp\Bigg\{-j\frac{4\pi f_{c}}{c}R_{t,i}\Bigg\}},
\end{equation} 
where $R_{t,i}$ is the range to the $i^{th}$ body part at time $t$, $f_{c}$ is the transmit center frequency, $c$ is the speed of light, and the amplitude $a_{i}$ is the square root of the power of the received signal as given by the radar range equation \cite{RichardsFRSP}. Thus, RF sensors provide a complex-time series of measurements in the form $x[t] = I[t] + jQ[t]$.

Typically, this data stream is re-shaped into a 2D matrix for each RF receive channel, so that the columns represent \textit{fast-time}, e.g. the analog-to-digital converter samples, and the rows represent \textit{slow-time}, e.g. pulse number.  The range ($R$) between the radar and any scattering point is found from the round-trip travel time ($t_{d}$) as
$    R=c t_{d}/2$.
In an FMCW radar system, the travel time can be found by mixing the received signal with the transmitted signal and filtering out high frequency components to obtain the beat frequency, $f_{b}=f_{t}-f_{r}$, which is the difference in the instantaneous frequencies of transmit and receive signals, $f_{t}$ and $f_{r}$, respectively.  Since the chirp rate is
$ \gamma=B / \tau = f_{b}/ t_{d}$, the range is found as
$    R=c \tau f_{b} / 2B$,
where $\tau$ is the pulse width.
The radial velocity of motion, $v_{r}$, is given by the Doppler shift,
$ f_{D} = 2 v_{r} f_{t} / c$,
which may be found by taking the Fast Fourier Transform (FFT) across pulses for a specific range bin.
The significance of these relations is that the range and velocity estimates obtained from RF sensors are independent measurements.

To enable fine grain motion recognition, it is also important for the radar to have sufficient resolution so as to distinguish the motion of the left and right hands, as well as fingers.  The minimum interval between two adjacent targets that can be discriminated by the radar in the radial direction is defined as the range resolution \cite{FMCW}, and is given by $\Delta r=c / 2B$.
The velocity resolution determines the minimum difference in velocity that can be perceived by the RF sensor, and, mathematically, is inversely proportional to the coherent processing interval or CPI, during which the target is illuminated. If $T_{f}$ is the CPI and $\lambda$ is the wavelength, then the velocity resolution \cite{FMCW}, $v_{res}$ is given by 
$   v_{res}=\lambda / 2T_{f} = \lambda / 2 N_{p} \tau$,
where $N_{p}$ is the number of pulses transmitted during a CPI.

With a bandwidth of $4 GHz$, center frequency of $77 GHz$ and a CPI of $40 ms$, the RF transmit waveform offers a range resolution of $0.0375m$  and a velocity resolution of $0.0487 m/s$. These numbers indicate that this sensor is capable of recognizing fine-grained motion characteristic of ASL.

\subsection{Experimental Design}
 The data were collected in a laboratory setting, where the sensor was placed on a table at an elevation of 0.91 m from the ground.  Participants sat on a chair directly facing a computer monitor, which was placed immediately behind the radar system. The monitor was used to relay prompts indicating the signs to be articulated. The radar system was positioned at a distance of 1.5 meters in front of the participant.

4 fluent ASL signers took part in the IRB-approved study, of whom 2 were Deaf and 2 were CODAs. The experiments included 100 ASL signs, as shown in Table 1, which were selected from the ASL-LEX2 \cite{ASL_LEX_2016} database to include signs of high frequency,
but not phonologically related to ensure a diverse dataset in terms of both handshapes and sign kinematics. The participants repeated each sign 5 times.  
The same experiment was repeated with 12 hearing participants, who did not know sign language. These participants were shown the signs prior to the experiment to familiarize them with the task.  During the experiment, immediately prior to recording, the participants were prompted with a video of each sign in isolation, and asked to repeat it.  Participants were presented with a random ordering of single signs minimize coarticulation during sign production. A total of 2000 fluent sign samples and 6000 imitation signing samples were collected.

\begin{figure}[t!]
\centering
\includegraphics[width=7cm]{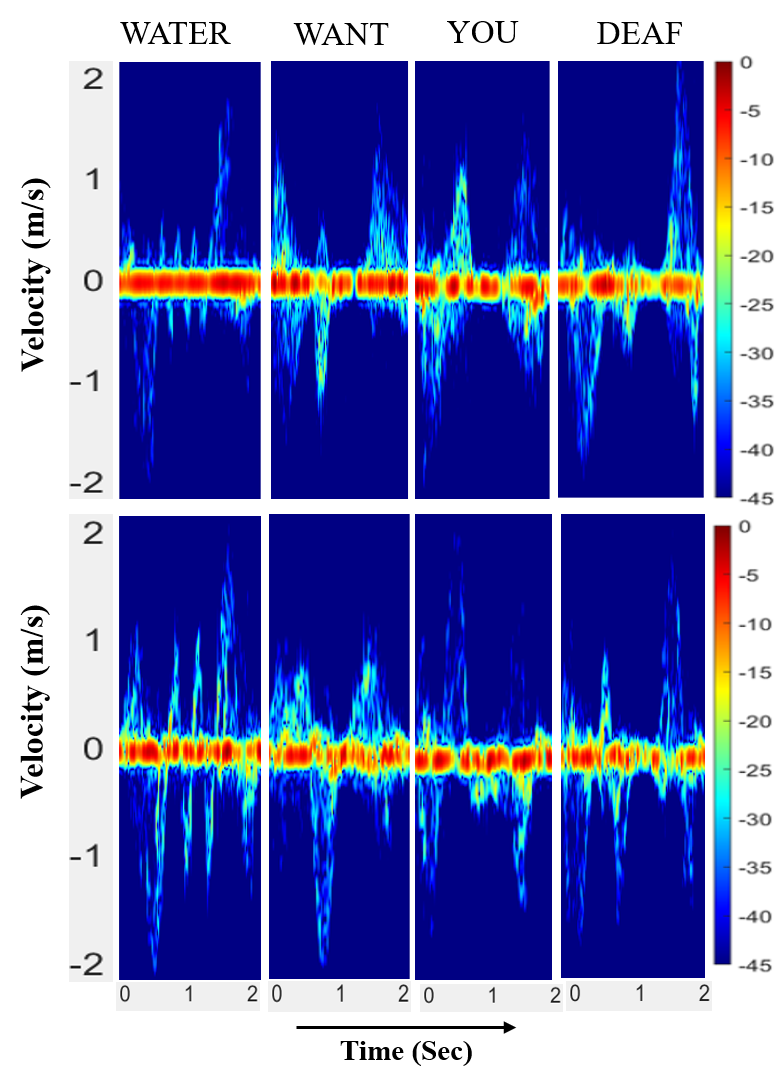}  \caption{Micro-Doppler signatures acquired from fluent (row 1) and imitation (row 2) signers.}
  \label{fig: Spectrograms}
\end{figure}

\subsection{RF Data Pre-Processing}
The kinematic properties of signing are captured by the frequency modulations in the phase of the received signal.  
Micro-motions \cite{ChenMD}, e.g. rotations and vibrations, result in micro-Doppler ($\mu D$) frequency modulations centered about the main Doppler shift, which is caused by translational motion. Signing results in a time-varying pattern of micro-Doppler frequencies. Each sign generates its own unique patterns, which can be revealed through time-frequency analysis.
The \textit{micro-Doppler signature}, or spectrogram, is found from the square modulus of the Short-Time Fourier Transform (STFT) of the continuous-time input signal $x(t)$   and can be expressed in terms of the window function, $h(t)$, as
\begin{equation}
    S(t,\omega)= \Big| \int^{\infty}_{-\infty} h(t-u) x(u) e^{-j \omega t} du \Big|^{2}.
\label{eq:stft}
\end{equation}

Ground clutter from stationary objects, such as furniture and the walls, will appear in the micro-Doppler signature as a band centered around 0 Hz.  At 77 GHz,  elimination of low-speed signal components during clutter filtering results in performance degradation \cite{9187644}, therefore no filtering was applied on the data.   Samples of the micro-Doppler signatures for both fluent and imitation ASL users are shown in Fig. \ref{fig: Spectrograms}.

\section{Estimation of Signing Kinematics}\label{sec:kinematics}
The most relevant kinematic information of a sign can be extracted from the motion associated with the arms and hands. Thus, the prosody of ASL is encoded in the velocity trace of the sign, since this is the simplest interpretation of the motion of the hands and arms of the signer \cite{Ronnie}.  In RF sensing, the micro-Doppler signature captures this velocity trace since the Doppler frequency is proportional to velocity.  In this section, we describe the processing steps taken to extract three kinematic properties of ASL:  hand speed, type of signs (one-handed vs. two-handed), and the number of directionally isolatable components of the sign with the motion toward the radar, termed strokes (including transitions toward and from initial and final handshapes).

\subsection{Hand Speed}
\begin{figure}[b!]
\centering
\includegraphics[height=4.5cm]{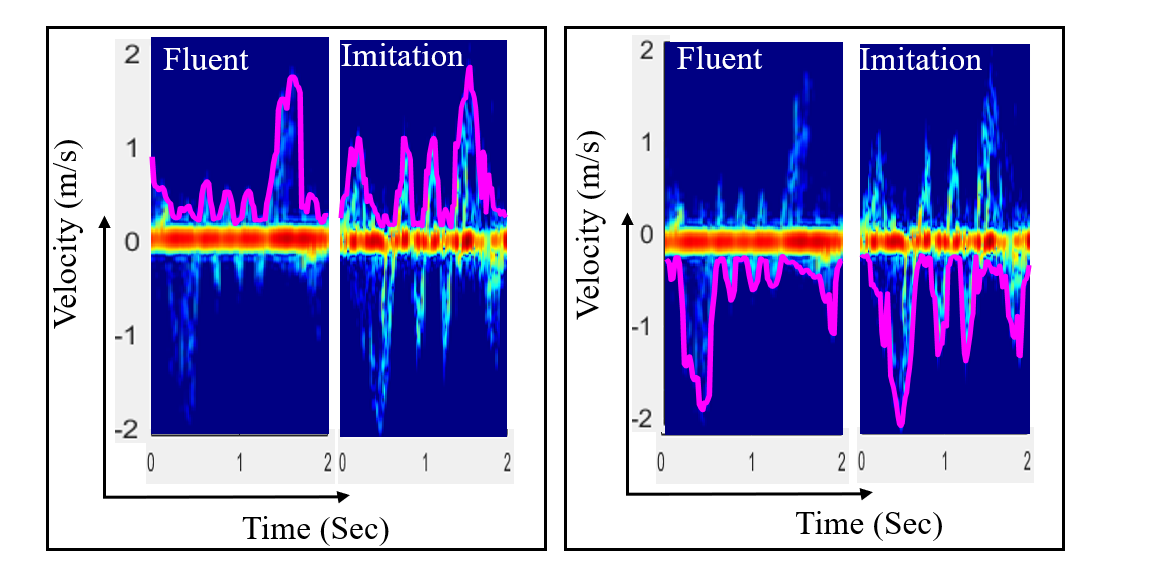} 
\caption{Spectrograms with upper and lower velocity envelope.}
\label{fig:spectrogram with upper and lower env}
\end{figure}

The speed of signing is measured by tracing the Doppler velocity of the upper and lower envelopes of the micro-Doppler signature. The upper envelope represents the radial velocity of the fastest point moving towards the radar, while the lower envelope gives the speed of the fastest point on the body moving away from the radar.  Thus, envelopes provide a means for learning the speed of the hands during signing. Envelopes are extracted by using an energy-based thresholding method \cite{Amin_SPM_Fall_2016}. First, the energy corresponding to each slow time index is computed, for which the first frequency bin whose corresponding
spectrogram amplitude is greater than or equal to a threshold is tagged as an envelope pixel.  The threshold is computed as the product of a pre-determined scaling factor and the energy at that slow-time index.
Figure \ref{fig:spectrogram with upper and lower env} shows an example of spectrograms for fluent and imitation signers, as well as their corresponding upper and lower envelopes. 

\begin{figure*}[t!]
\centering
\includegraphics[height=7.5cm]{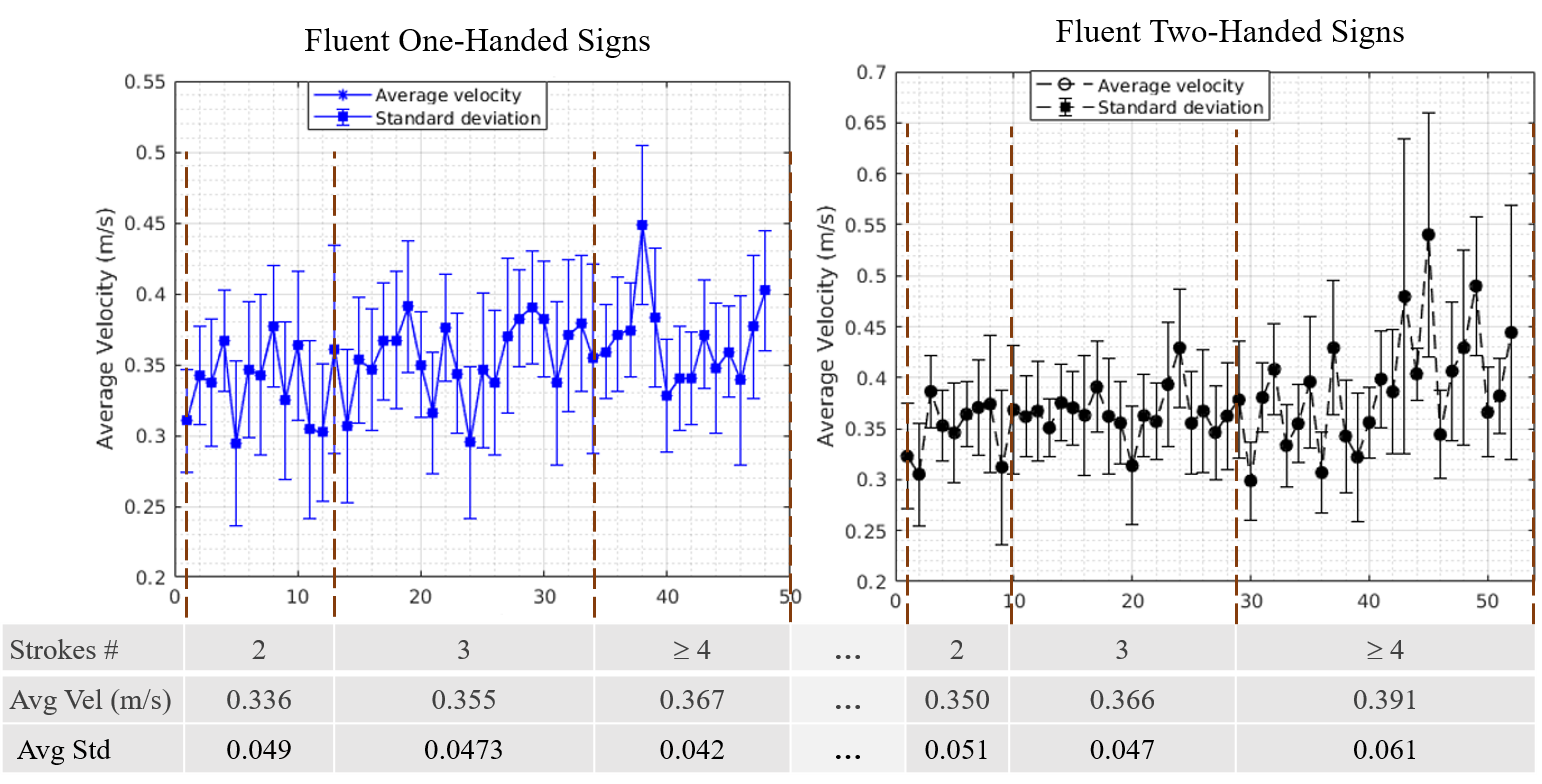} 
\caption{Average velocity with standard deviation for one-handed and two-handed fluent ASL signing.}
\label{fig: velocity with std fluent}
\end{figure*}

\begin{figure*}[t!]
\centering
\includegraphics[height=7.5cm]{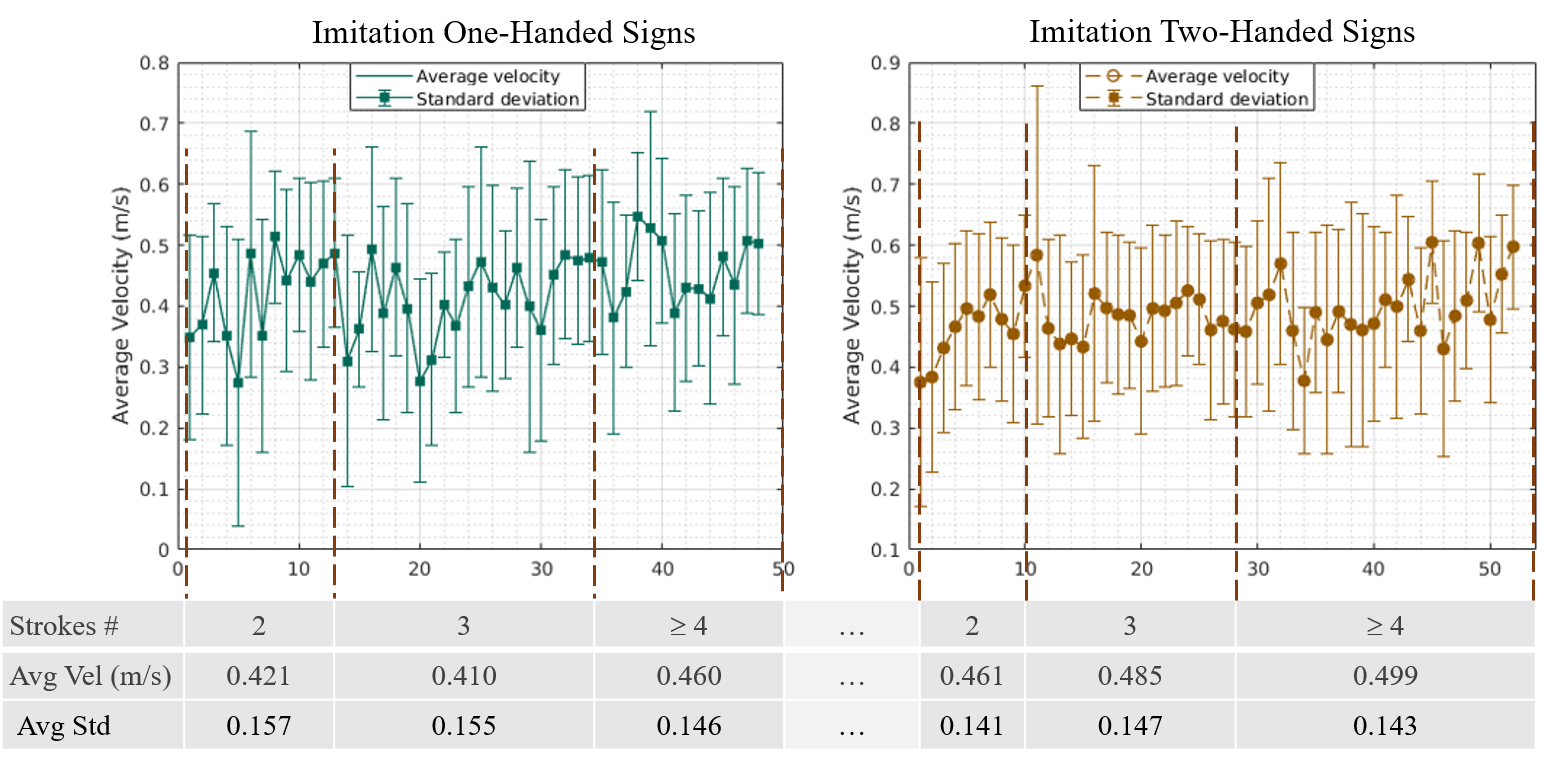} 
\caption{Average velocity with standard deviation for one-handed and two-handed imitation ASL signing.}
\label{fig: velocity with std imit}
\end{figure*}

The average hand speed during articulation of a sign is calculated by taking an average over each upper and lower envelope extracted from each spectrograms. Figure \ref{fig: velocity with std fluent} and \ref{fig: velocity with std imit} show the average hand speed and its standard deviation for fluent and imitations signers for 100 signs. The RF measurements show that the hand speed of imitation signers, on average (0.45 m/s), is greater than that of fluent signers (0.36 m/s), while the standard deviation is much greater.  This is reflective of the inconsistency between hearing participants in sign articulations.  The greater speed in hand movements of imitation signers may on the one hand seem surprising, as one might think someone less fluent would be more hesitant.  But, perhaps in part because hearing imitation signers perceive signing more akin to gesturing, than talking, their articulations are more rushed and sweeping.  In contrast, fluent signers articulate the sign within a tighter space, i.e. traverse less distance, but with calculated, precise expression.  This results in, on average, slower hand speeds.

Moreover, the RF measurements show that the average hand speed of two-handed signs are greater than that of one-handed signs. This may be in part because two-handed signs typically involve larger movements, while one-handed signs have finer-scale finger movements or hand shapes.

\subsection{Number of Strokes}
The number of strokes corresponds to the number of times the hands move towards the radar throughout the duration of the isolated sign (i.e. including transition to the initial handshape, and transition after the final handshape). In other words, number of positive peaks in the micro-Doppler signature correspond to the number of strokes. The number of strokes for a sign can be measured by applying peak detection algorithm to the upper envelope.  For repetitive motion (reduplicated signs), such as in signs \textsc{walk}, \textsc{water}, and \textsc{shop},  imitation signers were likely to err in production kinematics, producing an incorrect number of strokes (a typical error of early sign language learners).
From Figs. \ref{fig: velocity with std fluent} and \ref{fig: velocity with std imit}, it may be observed that as the number of strokes in a sign increases, hand speed also increases. 

\begin{figure}[t!]
\centering
\includegraphics[width=8.8cm]{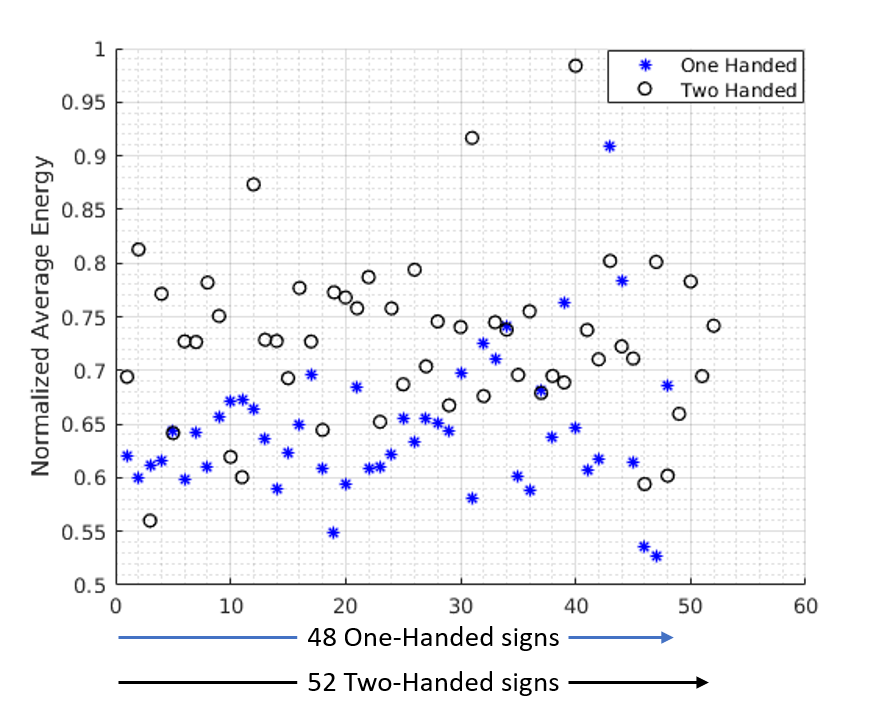} 
\caption{Energy-based sign type identification.}
\label{fig: energy one handed-two handed}
\end{figure}

\subsection{One-Handed vs. Two-Handed Signs}
Signs in ASL can be one-handed or two-handed.  One-handed signs result in less backscatter than two-handed ones.  Thus, the average received signal energy for a sign can be indicative of whether the sign involves one or two hands.  The total energy of a spectrogram is computed by summing the energy corresponding to each slow-time index.
This process is repeated for each spectrogram and then divided by the number of samples to find the algebraic mean. In this way, the total average energy for all 100 words is calculated. Figure \ref{fig: energy one handed-two handed} compares the average total energy for one-handed and two-handed signs. Note that energy of two-handed signs is distinctly higher than that of one-handed signs.  Thus, a threshold can then be designated for categorizing whether a sign is one-handed or two-handed.  we found that a threshold of 0.674 yielded a classification accuracy of 81\% for one-handed versus two-handed signs. 

The kinematic errors of real imitation signers can be quantitatively compared with that of fluent signers using the metrics of average speed ($V_{h}$), number of strokes ($N_{str}$), and handedness detection ($N_{h12}$), as shown in Table \ref{fig:real_stats}. Notice that the deviation in average speeds of imitation signers is greater than that of fluent signers. In addition, the errors in repetitions during the articulation of signs, as indicated by the number of strokes, is also significantly higher for imitation signers.  This is consistent with the visual observations of fluent and hearing participants during experiments.

\begin{table}[h!]
\centering
\caption{Statistics of real fluent and imitations signer data.}
\includegraphics[height=2.2cm]{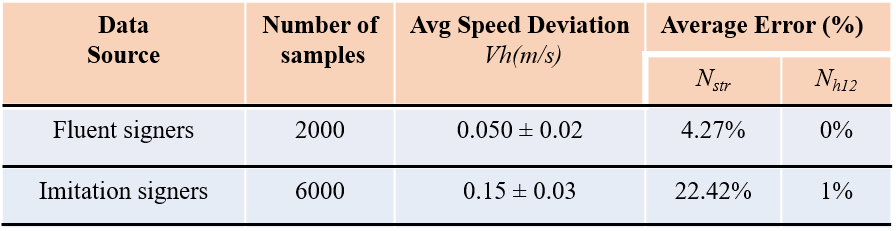} 
 \label{fig:real_stats}
\end{table}

\begin{figure*}[t!]
\centering
\includegraphics[width=0.95\textwidth]{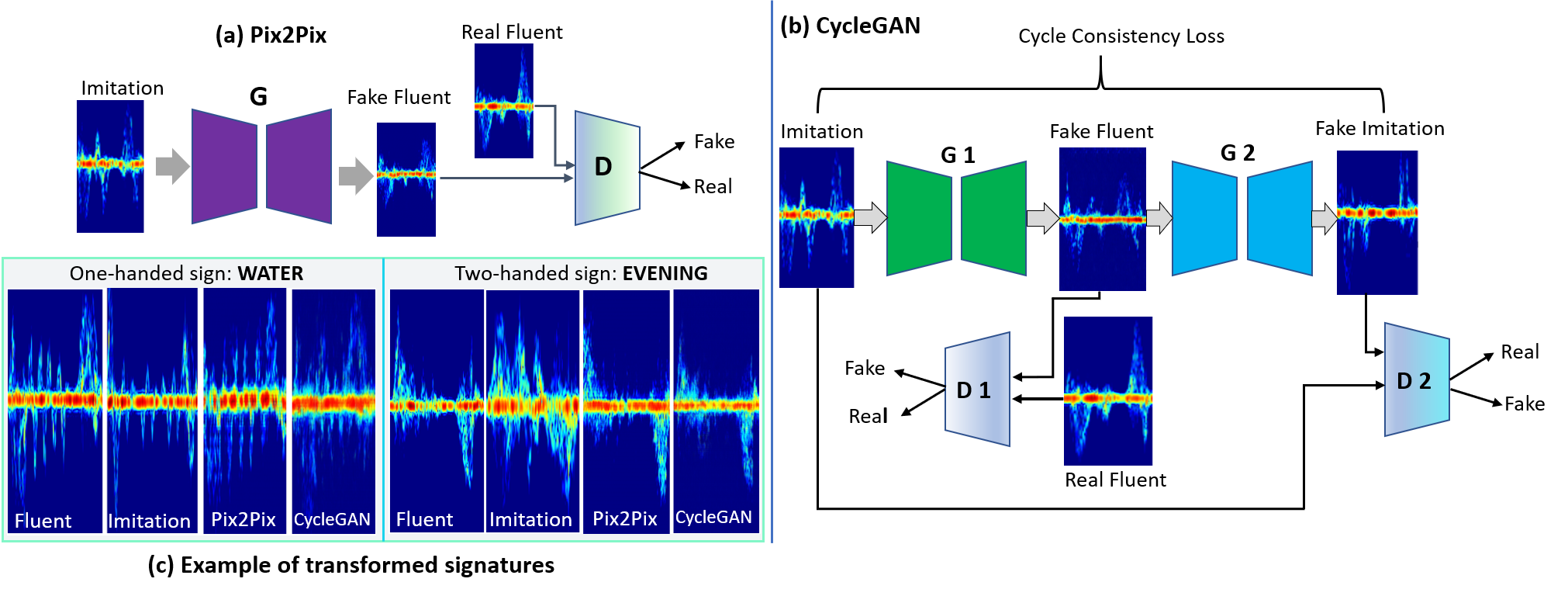}
\caption{ Imitation to fluent sign transformation using (a) Pix2Pix  and (b) CycleGAN. (c) Examples of transformed signatures.}
 \label{fig: pix2pix and cyclegan}
\end{figure*}

\section{Adversarial Learning Approaches}\label{sec:adversarialLearning}
\label{sec:Adversarial Learning Approaches}
Compilation of large datasets for training state-of-the-art DNNs is difficult when human subjects are involved, due to the time spent in measuring numerous iterations of each class. In previous work \cite{9187644}, ASL recognition using conventional supervised machine learning was explored due to the small amount of available real data: 9 samples per class per sensor, using a total of 5 sensors.  The minimum-redundancy maximum-relevance (mRMR) method was used to select 150 handcrafted features for input to a random forest classifier, resulting in a classification accuracy of 72.5\% for 20 ASL signs.
Later, a slightly larger dataset was acquired (on average of 40 samples per class per sensor for 3 sensors at different frequencies) to train a DNN for fusion of multi-frequency sensor data to achieve an accuracy of 95\% for the same 20 ASL signs \cite{Gurbuz_multi_freq}.  The limitations in the amount of available real training data also limited the depth and accuracy of the DNNs utilized.

One approach that has been used in some studies \cite{DeepASL2017,SignFi2018} is to instead use imitation signing data for both training and testing of algorithms.  However, this can lead to over-optimistic results \cite{Gurbuz_multi_freq} due to the differences in production between imitation and fluent signers, which are also captured by the RF sensor measurements as presented in Section III. This is further evidenced by the ability to distinguish between the RF data from fluent versus imitation signers using a support vector machine classifier \cite{ASLR_JSens2020}. Thus, we wish to emphasize that in this work, all DNNs have been tested on ASL signs articulated only by fluent signers.

The question of whether imitation signing data can be leveraged in any way to train DNNs for ASL recognition of fluent signers is an interesting avenue to explore.  Due to the differences in data distribution, direct use of imitation signing data as training data is not effective: when a convolutional neural network (CNN) is trained on imitation signing data and tested on fluent ASL-R dataset, only 24\% accuracy is attained \cite{Gurbuz_multi_freq}.  One possible remedy could be to use domain adaptation techniques to transform imitation signing data into signatures that better match the distribution of fluent ASL data, as discussed next.

\subsection{Transformation of Imitation Signing Signatures}
Image translation is a class of computer vision techniques where the goal is to learn a mapping between an input and an output image. A number of image-to-image translation techniques such as Pix2Pix\cite{Isola2017ImagetoImageTW}, CycleGAN\cite{Zhu2017UnpairedIT} and TravelGAN\cite{Amodio2019TraVeLGANIT} have been proposed in the literature.  As CycleGAN has been shown to outperform TravelGAN on RF signatures \cite{Rahman_RadarConf2021}, in this work we consider the efficacy of both Pix2Pix and CycleGAN for transformation of imitation signing data. The architectures of both techniques are illustrated in Figure \ref{fig: pix2pix and cyclegan}.

\subsubsection{Pix2Pix} Pix2Pix is a type of conditional GAN (cGAN), where the generation of the output image is conditioned on the input; in this case, a source image. The generator of Pix2Pix uses the U-Net \cite{ronneberger2015unet} architecture. In general, image synthesis architectures take in a random vector as  input, project it onto a much higher dimensional vector via a fully connected layer, reshape it, and then apply a series of de-convolutional operations until the desired spatial resolution is achieved. In contrast, the generator of Pix2Pix resembles an auto-encoder. The generator takes in the image to be translated, compresses it into a low-dimensional vector representation, and then learns how to upsample it into the output image.
The generator is trained via adversarial loss, which encourages it to generate plausible images in the target domain. The generator is also updated via an $\ell_1$-loss measured between the generated image and the expected output image. This additional loss encourages the generator model to create plausible translations of the source image.

The architecture of the  discriminator is a PatchGAN / Markovian discriminator \cite{patchgan} that works by classifying individual ($N \times N$) patches in the image as “real vs. fake,” as opposed to classifying the entire image. This enforces more constraints that encourage sharp high-frequency detail in the output images.
The discriminator is provided both with a source image (in this case, an imitation signing signature) and the target image (fluent signing signature) and must determine whether the target is a plausible transformation of the source image.

One limitation of Pix2Pix is that since it is a paired image-to-image translation method, the total number of synthetic samples generated is identical to the number of real imitation signing signatures acquired. In this work, a total of 6,000 transformed signatures are synthesized using Pix2Pix.

\subsubsection{CycleGAN} In constrast to Pix2Pix, CycleGAN  is a GAN for unpaired image-to-image translation.  Thus, a greater amount of synthetic data can be generated than the real imitation samples used at the input of the network. For two domains $A$ and $B$, CycleGAN learns two mappings: \textit{\textbf{$G$:$A$\textrightarrow$B$}} and \textit{\textbf{$F$:$B$\textrightarrow$A$}}. CycleGAN translates an image from a source domain A to a target domain B by forming a series connection between two GANs to form a “cycle”: the first GAN tries to synthesize “fake fluent” ASL data from the imitation signing data, while the second GAN works to reconstruct the original sample, synthesizing “fake imitation” ASL samples. Thus, the network tries to minimize the cycle consistency loss, i.e. the difference between the input of the first GAN and the output of second GAN.

Each CycleGAN generator is comprised of three sections: an encoder, a transformer, and a decoder. The input image is fed directly into the encoder, which shrinks the representation size while increasing the number of channels. The encoder is composed of three convolution layers. The resulting activation is passed to the transformer, a series of six residual blocks. It is then expanded again by the decoder, which uses two transpose convolutions to enlarge the representation size, and one output layer to produce the final transformed image.
The discriminators are comprised of PatchGANs - fully convolutional neural networks that look at a “patch” of the input image, and output the probability of the patch being “real.” This is both more computationally efficient than trying to look at the entire input image, and is also more effective since it allows the discriminator to focus on more localized features, like texture. 

\subsubsection{Comparison of Pix2Pix and CycleGAN}
Samples of Pix2Pix and CycleGAN transformed signatures are shown in Figure \ref{fig: pix2pix and cyclegan} (c).  Although the general trends in the signatures are consistent, the Pix2Pix signatures have greater visual resemblance to the signature from fluent ASL users.  CycleGAN signatures appear more faded and blurry, especially in regions outside the 0 Hz ground clutter returns.   These differences can be quantitatively compared via the kinematic properties of ASL, which can be extracted from RF data as described in Section \ref{sec:kinematics}.  Table \ref{fig:kincomp} lists the mean error and standard deviation of hand speed, $V_{h}$, as well as the percentage of erroneous samples of strokes, $N_{str}$, and handedness detections, $N_{h}$, for the number of synthetic samples, $N_{s}$.  While Pix2Pix can only transform 6,000 samples, CycleGAN is used to generate both 6,000 and 50,000 samples.

\begin{table}[b!]
\centering
\caption{Comparison of Kinematic Errors in Pix2Pix and CycleGAN Signatures.}
\includegraphics[height=2.7cm]{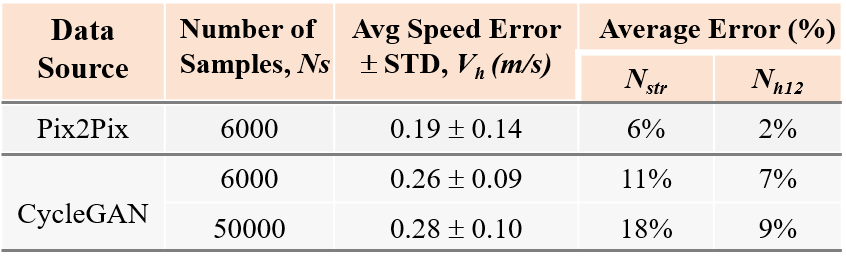} 
\label{fig:kincomp}
\end{table}

Pix2Pix signatures show better adherence to the kinematic properties of fluent signing than CycleGAN.
Notice that on average, the CycleGAN signatures exhibit more error in hand speed, number of strokes and detection of handedness relative to those generated by Pix2Pix.  Increasing the number of generated CycleGAN samples has only a slight detrimental effect on hand speed, while more significant errors are induced in the number of strokes and handedness. 


Note that there are two different causes for kinematic errors:  first, the lack of fluency in the language, and second, the network itself.  Let us first consider the reasons for why Pix2Pix significantly outperforms CycleGAN.  In prior work, we showed that the generative process creates synthetic data with significant kinematic errors  \cite{Erol_TAES_2020}, which are described more in the next section. Networks generate kinematic errors because RF data is not naturally an image, but converted into a 2D format via time-frequency analysis (Section II-C).  Hence, spatial correlations are not based on physical proximity (as in optical images), but on the distribution of velocity across the human body and the constraints imposed by the skeleton.  However, the GAN architectures are not supplied with any information or metric pertaining to these constraints, resulting in synthetic samples that bear spatial resemblance, but in fact correspond to physically impossible movement.  The CycleGAN architecture includes two generators, in contrast to the single generator of Pix2Pix; hence, the greater the amount of kinematic errors exhibited in the CycleGAN synthesized samples.  

Moreover, imitation signing data itself has significantly more error in average speed as well as the number of strokes, as was shown in Table \ref{fig:real_stats}.  That these errors persist in the domain adapted signatures can be seen by observing that the average errors in Pix2Pix and CycleGAN synthesized data remain significantly greater than the levels observed in fluent signing data. In fact, the error in average speed of Pix2Pix data exceeds that observed even in real imitation signing data.

\subsection{Direct Synthesis of ASL Sign Signatures}
An alternative to transformation of imitation signatures is to instead use a small amount of real, fluent ASL data as input to a GAN, which generates a larger number of synthetic samples for training.  
In our prior work \cite{Erol_TAES_2020,Erol_MBGAN_2020}, several different types of architectures have been explored for synthetic data synthesis, including auxiliary-conditional GANs (ACGANs), conditional variational autoencoders (CVAE) and WGANs, but all were found to generate data that exhibits significant discrepancies from that of real RF signatures.  Examples include
\begin{itemize}
    \item \textit{Disjoint components micro-Doppler:}  Real micro-Doppler signatures are connected and continuous, because all points on the human body are connected with each other, forming a continuous spread of velocities.  This prevents human RF signatures from having disjoint components or regions in the signature.  
    \item \textit{Leakage between target and non-target components:} A benefit of GANs is that sensor-artifacts can also be synthesized, but sometimes this results in leakage (connected segments) between target movements and sensor artifacts or noise, which are not physically possible.
    \item \textit{Incorrect shape of signature:}  When the shape of the micro-Doppler is distorted, with additional peaks, or symmetric reflections about the x-axis, these components correspond to physically impossible movements; e.g., a person whose hand simultaneous moves towards and away from the radar, additional repetitions, or sudden motion back and forth that are not normally part of the sign. 
\end{itemize}
While these erroneous components may not seem significant visually,   they ultimately correspond to kinematically impossible articulations, which, when used as training data,  incorrectly trains the DNN and significantly degrades classification accuracy.

\begin{figure}[t!]
\centering
\includegraphics[height=11.8cm]{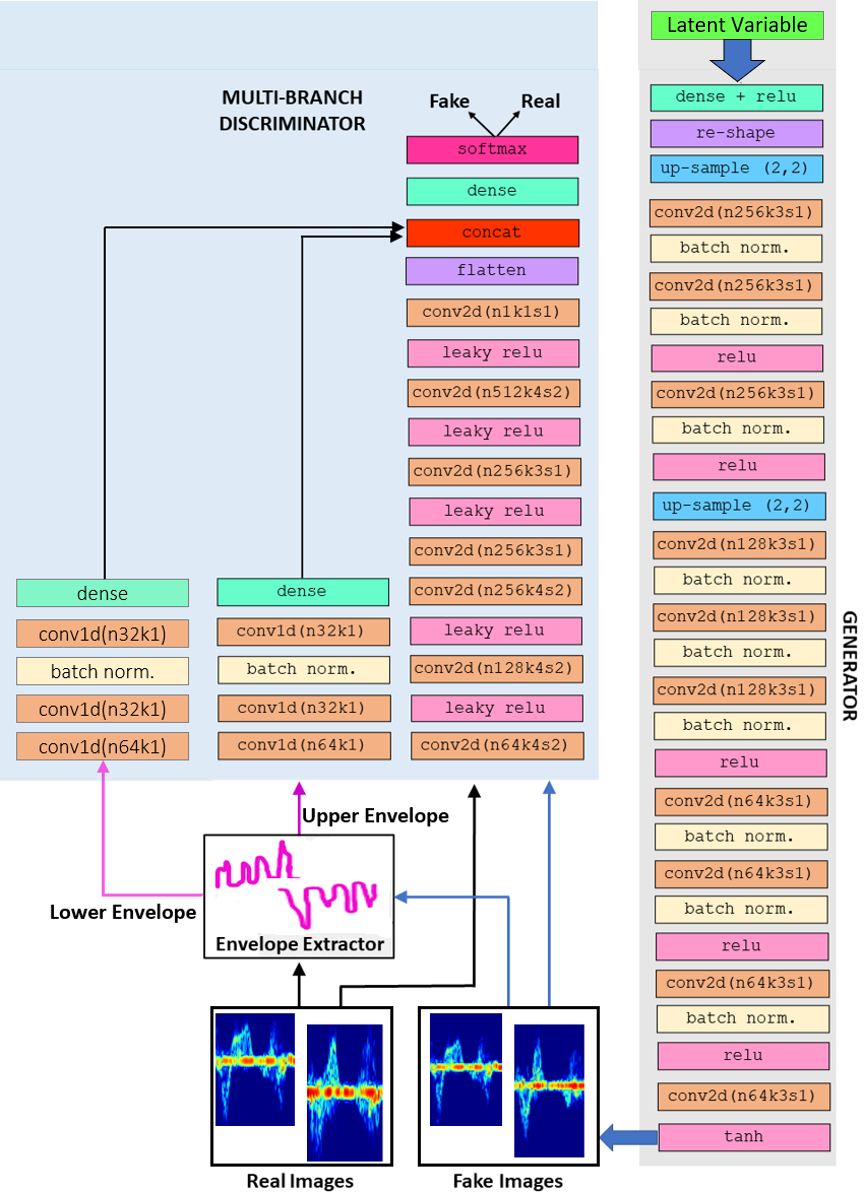}
\caption{Proposed 3-branch discriminator MBGAN.}
 \label{fig: MBGAN archi}
\end{figure}

One way to mitigate such problems is to design the GAN so as to enable greater emphasis on preservation of the shape of the envelope.  The envelopes correspond to the maximum velocity towards/away from the radar; so, from the standpoint of hand kinematics, the synthetic signatures should conform to, and not exceed the envelope profiles of source data.  In prior work \cite{Erol_MBGAN_2020,Rahman_LRMBGAN_2021}, a multi-branch GAN (MBGAN) architecture with an additional auxiliary branch in the WGAN discriminator, which took as input the upper envelope, was proposed as a means of ensuring kinematic accuracy when synthesizing micro-Doppler signatures of different ambulatory gaits, such as walking, limping, or taking short steps. However, during production of sign language, the hands may move towards or away from the radar, so both the upper and lower envelopes are important for maintaining critical kinematic features. Hence, in this work, we incorporated two additional auxiliary branches in the discriminator:  one that takes the upper envelope as input, and a second that takes the lower envelope as input.  The resulting MBGAN with 3-branch discriminator is shown in Figure \ref{fig: MBGAN archi}.  The generator is comprised of 10 convolutional layers; each layer is followed by batch normalization
with 0.9 momentum and a Rectified Linear Unit (ReLU) activation function. The main branch of the discriminator is
an 8-layer CNN, where each layer is followed by a Leaky-ReLU activation function. Each auxiliary branch is comprised of three 1D-convolutional layers.
The outputs of the dense layers are concatenated with the flattened output of the main discriminator.

The kinematic errors incurred in the synthetic signatures generated by WGAN and MBGAN are compared in Table \ref{fig:wmbgancomp}.
The both networks used 75\% of the fluent signing data as input during training, and the remaining are used for testing.
Notice that the synthetic training data generated by WGAN or MBGAN outperform both Pix2Pix and CycleGAN with respect to generating signatures that have greater kinematic fidelity to fluent signing data.  While the kinematic errors in WGAN generated signatures increase as the network generates a greater amount of synthetic samples, the errors in MBGAN signatures are  fewer, and constant over sample size - only a slight drop in the accuracy in replicating the correct number of strokes is incurred, from \%99.5 to \%98.  Visually, MBGAN signatures may be observed to have greater resemblance to fluent ASL samples in comparison with the WGAN samples, as shown in Figure \ref{fig:spect_gan}.  Note that in comparison, peaks in WGAN signatures are not as clearly constructed, slightly faded, and have envelopes whose shape has some differences from the envelope of the fluent ASL signature.

\begin{table}[t!]
\centering
\caption{Comparison of Kinematic Errors in WGAN and MBGAN Signatures.}
\includegraphics[height=3.1cm]{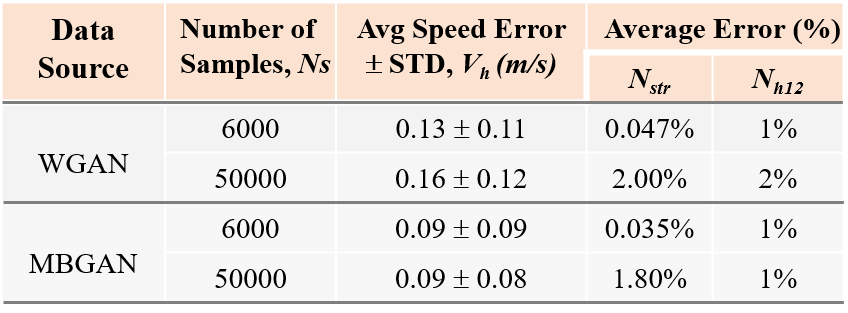} 
\label{fig:wmbgancomp}
\end{table}

Signatures with kinematic errors may divert the classifier in the wrong direction during the feature extraction, and thereby result in poor recognition performance. Hence, it is important to identify and exclude the incorrect kinematic signatures generated in GANs synthesis. In the next section, several kinematic rules are defined and the synthesized data are sifted by these constraints.


\subsection{Kinematic Sifting}
Although the 3-branch discriminator MBGAN does have the intended effect of generating signatures with greater kinematic fidelity to fluent ASL, relative to the other networks considered, it is possible that it still generates kinematically unrealistic synthetic samples.  Ideally, we wish to generate training data that is statistically independent, diverse, and representative of the range of potential variations within possible articulations of each sign. The presence of kinematically erroneous samples can have a corrupting effect that leads to confusion between different signs.  Thus, we seek to remove such samples.

One way of removing outliers is to determine a boundary in feature space based on the measured, fluent ASL data acquired.  Using Principal Component Analysis (PCA) or t-SNE \cite{tsne}, each sample can be projected to a 3-D feature space and a convex hull encompassing all samples computed.  The convex hull thus forms a boundary; any synthetic samples lying beyond this boundary could be excluded from the training dataset as ``erroneous." However, it is still possible for samples within the convex hull boundary to be kinematically flawed.  Instead of relying on the PCA-based convex hull, instead we identify and sift flawed synthetic data based on the following kinematic properties:

\begin{figure}[t!]
\centering
\includegraphics[width=7.8cm]{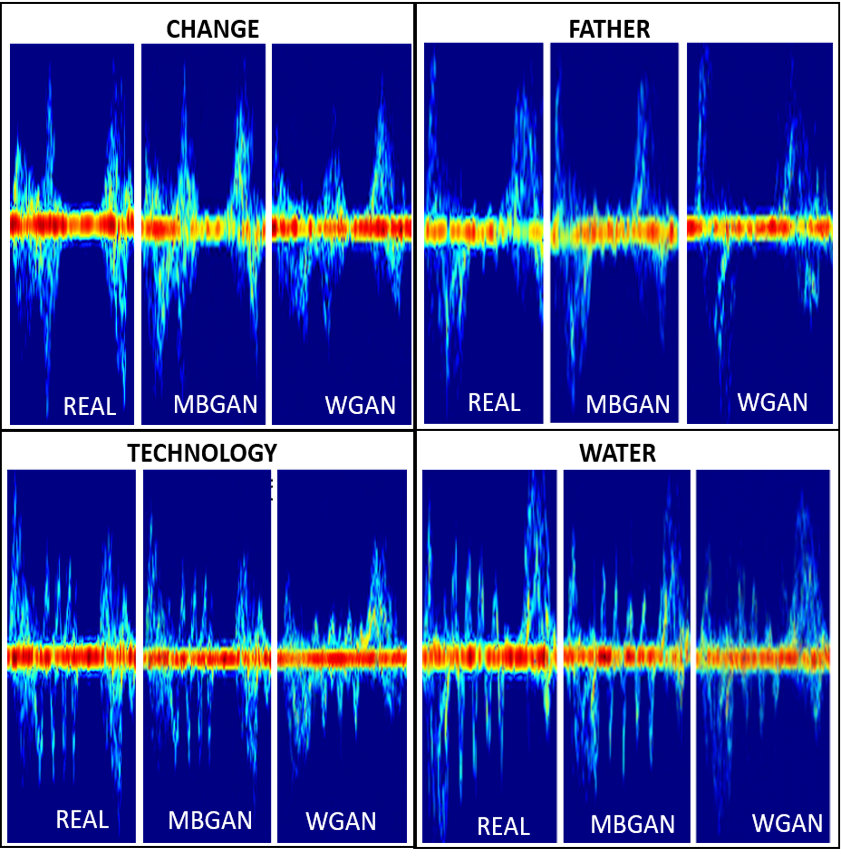}
\caption{Example of WGAN and MBGAN generated spectrograms.}
 \label{fig:spect_gan}
\end{figure}


\begin{itemize}
    \item {\textbf{\textit{Rule 1 - Number of strokes:}}} 
    The number of strokes estimated using the peak detection algorithm described in Section \ref{sec:kinematics} is compared with the number of strokes listed for each sign in Table \ref{fig: 100 ASL word}, as given by ASL-LEX. If the detected number of strokes for a synthetically generated signature is incorrect, then this synthetic sample is removed from the dataset.   For example, this rule will preclude signatures corresponding to a signer utilizing an incorrect number of repetitions.
    \item {\textbf{\textit{Rule 2 - Total Energy:}}} As energy is related to whether a sign is one handed or two handed, ensuring the synthetic data lies within reasonable energy bounds, given that we know whether the sign is one or two handed, can be an effective criterion.  The rule is tested by first finding the average total energy and its standard deviation from the real, fluent ASL data.  Then, for each synthetic signature, the total energy is calculated and checked to see whether it falls within $\pm$ $1$ standard deviation of the average total energy of the real signatures.  This is tested on a class-by-class basis. If the criterion holds true then the sample is regarded as kinematically valid, otherwise it is sifted out.
    
    \item {\textbf{\textit{Rule 3 - Envelope Matching:}}} The envelope of the spectograms is a time-series curve and the similarity between curves can be measured by taking into account both the location and ordering of the points along the curve \cite{Amr_motion}.  Dynamic Time Warping (DTW) \cite{DTW2} is a commonly used curve matching technique that measures the similarity between two temporal sequences. To apply envelope matching as a kinematic metric , first, for each class, the average DTW distance and standard deviation are calculated from the combinations of all real samples. Then the DTW distance for each synthetic samples are computed with respect to each real samples, from which the average distance is found. Then this average distance is examined whether it falls within $\pm$1 Standard deviation of the average DTW distance for that class. If it is within the limit then the sample is kinematically valid; otherwise sifted out as kinematically invalid.

\end{itemize}

\begin{figure}[t!]
\centering
\includegraphics[width=7.2cm]{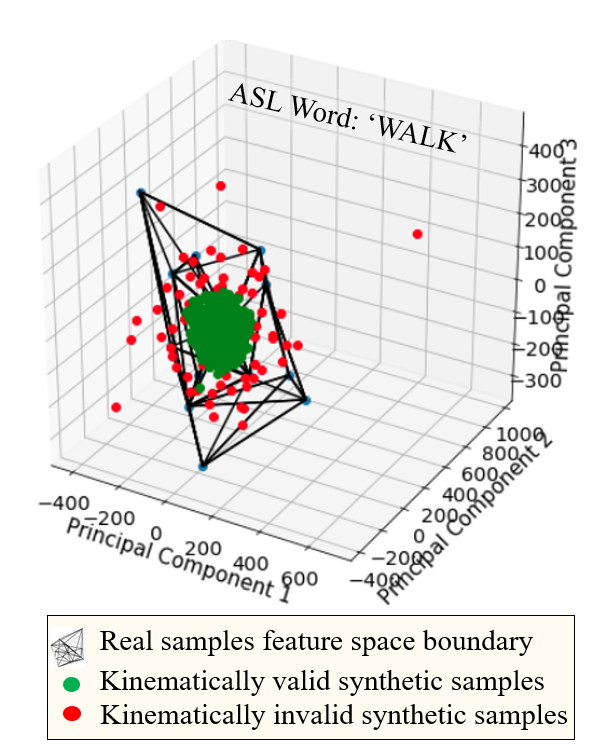}
\caption{Kinematically  sifted synthetic samples projected on real sample's feature space boundary.}
 \label{fig:statistic_vs_kinematic}
\end{figure}

Kinematic sifting provides for tighter constraints than the convex hull boundary.  Consider synthetic samples for the ASL sign \textsc{walk} projected to a 3D space using t-SNE, as shown in Figure \ref{fig:statistic_vs_kinematic}. The boundary found based on the convex hull derived from the real fluent ASL samples is shown with solid black lines.  The synthetic samples are shown as dots.  Notice that most fall within the convex hull boundary, while some are outliers.  With statistical sifting, only the outliers outside this boundary would be sifted out of the training dataset.  However, if we apply the kinematic rules described above, we may see that there are many kinematically invalid samples (shown in red) that remain within the hull.  The valid samples (shown in green) form a tight nucleus within the convex hull.  Hence, the kinematic rules form a more stringent constraint.  


Table \ref{fig:siftedncomp} provides a listing of the number of synthetic samples sifted out, $N_{sft}$, by kinematic rules for the Pix2Pix, CycleGAN, WGAN, and MBGAN networks, and the kinematic errors based on the sifted synthetic datasets.  For all synthesis methods, comparison with error metrics reported in Tables II and III
shows that the sifting process reduced the average error in the remaining data.  As the number of samples generated increases, kinematic errors only slightly increase.  CycleGAN appeared to be the network most prone to errors, with the greatest number of samples failing the kinematic rules, and, hence, was excluded from the final synthetic dataset. In contrast, even after sifting, the proposed MBGAN remains the network that results in synthetic signatures that exhibit the greatest kinematic fidelity to fluent signing data.

\begin{table}[t!]
\centering
\caption{Comparison of Kinematic Errors in Signatures After Sifting.}
\includegraphics[height=3.9cm]{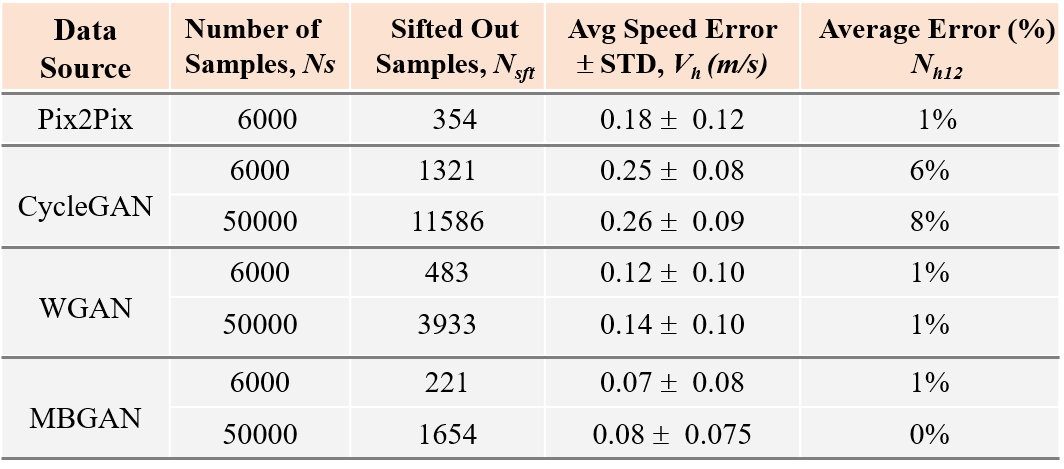} 
\label{fig:siftedncomp}
\end{table}

\section{ASL Recognition Results}\label{sec:ASLRecognition}
In this section, the resulting classification accuracies obtained using the various methods for synthesizing training data are compared.
\subsection{DNN Architectures}
In previous work \cite{Seyfioglu_GSRL_2017}, Convolutional Auto-encoders (CAEs) were shown to be effective when small, yet reasonable, amounts of real data are available for training, outperforming  transfer learning from weights pre-trained using ImageNet \cite{imagenet} for VGG\cite{Vgg} and Resnet \cite{Resnet}.  Consequently, in this work, a four-layer convolutional autoencoder (CAE) has been utilized to classify the 100-sign fluent ASL dataset. 
CAEs use unsupervised pre-training to initialize the network near a good local minima. In each layer, a filter concatenation technique is employed, in which a filter size of $3 \times 3$ and $9 \times 9$ were concatenated to take advantage of multilevel feature extraction. After training the CAE model, the decoder was removed, and two fully connected
layers with 256 neurons followed by a dropout of 0.55 were added after flattening the output of the encoder. At the output, a softmax layer with 100 nodes was employed for classification.  During training, an ADAM \cite{adam} optimizer was utilized, along with a batch size of 16, learning rate of 0.0005 and 30 epochs. 

\subsection{Classification Accuracy}
\begin{figure}[t!]
\centering
\includegraphics[height=4.5cm]{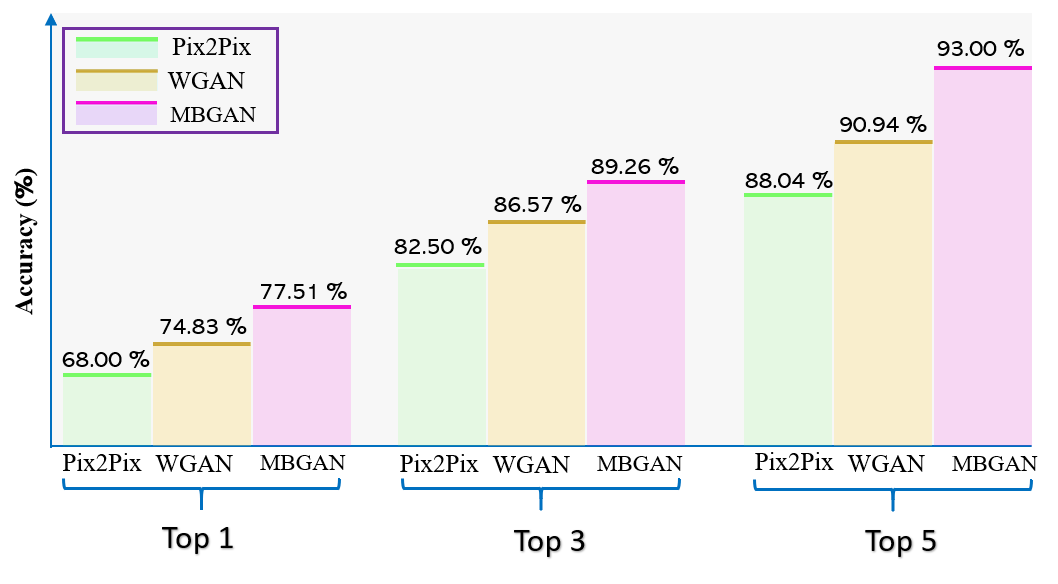}
\caption{Comparison of accuracies attained with Pix2Pix, WGAN, and proposed MBGAN methods for synthesizing training data. }
 \label{fig: BarFinal}
\end{figure}

The classification accuracies obtained using the CAE trained on the various sources of synthetic data are compared in Table \ref{fig:cae}, while the best performing techniques are compared in Figure \ref{fig: BarFinal} based on the Top-1, Top-3, and Top-5 accuracies.  The proposed approach of direct training data synthesis with MBGAN surpasses other conventional approaches by achieving a 77\% top-1 accuracy, 89\% top-3 accuracy, and 93\% top-5 accuracy.

\begin{table}[h!]
\centering
\caption{100 ASL Signs Recognition Using CAE.}
\includegraphics[width=8.6cm]{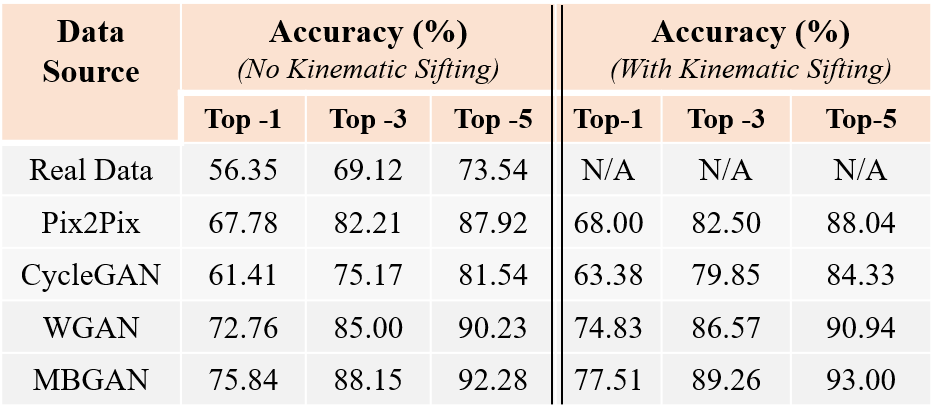}
\label{fig:cae}
\end{table}

\subsection{Implications and Discussion}
The most important conclusion we may draw from these results is based on the observation that the resulting classification accuracies are inversely related to the amount of kinematic errors in the synthetic data.  The greater the error, the lower the classification accuracy.  For all methods, sifting out samples that fail the kinematic rules results in performance improvement.  

As mentioned earlier in the paper, there are two sources of errors: namely, kinematic errors generated by DNNs used for adaptation and synthesis, and kinematic errors inherent to the original source data.  The direct synthesis approach with GANs have the benefit of utilizing fluent signer data in the synthesis process.  In contrast, the data synthesized via domain adaptation contains both sources of errors.
Note that the error in average speed reflected in Pix2Pix synthesized samples is $0.19 m/s$, which is greater than that computed from  the real data from imitation signers ($0.09 m/s$), Pix2Pix's source data.  In other words, Pix2Pix cannot compensate for the imitation signing errors, and exhibits additional model-based errors as well.

In comparing Pix2Pix results with that of WGAN, readers should be reminded that domain adaptation methods predominantly utilize a PatchGAN architecture in the discriminator, which operates on localized patches in the image, while WGAN and MBGAN discriminators operate on the entire image.  Both results in the literature \cite{Isola2017ImagetoImageTW} and comparisons we conducted on radar micro-Doppler signatures reveal that operations on patches are more effective than that on the entire image.  For example, Pix2Pix with a PatchGAN discriminator generates much crisper and textured synthetic signatures than Pix2Pix with a discriminator operating on the entire image.  This is likely because discriminators operating on the entire image cannot model the sharpness of high frequency components in the image as effectively.  Modeling high frequencies requires restricting attention to the structure in local image patches through the application of penalties at a patch-scale.  Despite utilization of the entire image, rather than patches, in the discriminator, WGAN synthesized signatures exhibited fewer kinematic errors than the domain adaptation networks we considered.  
 
 Although the error in average speed in WGAN signatures ($0.13 m/s$) is lower than that of Pix2Pix, it is greater than that of imitation signing.  These errors are due to the generation process itself, and can be mitigated through modification of GAN architecture, such as done in the proposed MBGAN.  In future work, we plan to explore extensions of the proposed approach (e.g. modifications of GAN architecture and inclusion of envelopes as an auxiliary input, as well as modifications to the loss function to include physics-based loss regularization \cite{Rahman_LRMBGAN_2021}) to adversarial domain adaptation to improve the resulting classification accuracy when imitation signing data is leveraged for model training.
 
 Another important open question for future work that relates to Explainable AI is to better understand the physical interpretation, i.e. underlying kinematic model, and nature of the diversity seen in GAN-synthesized micro-Doppler signatures.  For example, do the variations between synthetic samples correspond to plausible variations within a certain subject profile (physical or linguistic), or span all probable articulations within a class?  Improvements to the generation of synthetic data for training will require a better understanding of not just the statistical properties, but also the physical and linguistic properties of the synthetic samples to ensure good model generalization.


\section{Conclusion}
Although imitation signing has been used in some studies of sign language recognition, imitation signers exhibit significant differences in kinematics of sign production as compared with fluent signers.  This results in substantial statistical differences between imitation and fluent ASL data, which has rendered imitation data ineffective when used to train DNNs for fluent signing recognition \cite{Gurbuz_multi_freq}.  This work investigates the use of domain adaptation to transform imitation signing samples to have greater resemblance to fluent signing data, and compares the efficacy of this approach with direct generation of synthetic data from fluent signing data.  A novel approach to synthetic RF signature generation is proposed, which is shown to generate samples with greater kinematic fidelity than conventional GANs for transformation of imitation signing samples.  Proposed kinematic metrics are extracted from RF ASL signatures and used to evaluate GAN-generated synthetic data from a kinematic perspective.  The classification results obtained using a CAE were found to be directly proportionate to the kinematic fidelity of the synthetic data.  The proposed methods were used to achieve 77\% top-1 accuracy, 89\% top-3 accuracy, and 93\% top-5 accuracy for the recognition of 100 ASL signs.

\section*{Acknowledgment}
The   authors   would   like   to   thank   Dr.   Caroline   Kobek-Pezzarossi  from  Gallaudet  University,  Washington  D.C.  and Dr.  Dennis  Gilliam  from  AIDB  for  their  support  of  this research.

\begin{IEEEbiography}[{\includegraphics[width=1in,height=1.25in,clip,keepaspectratio]{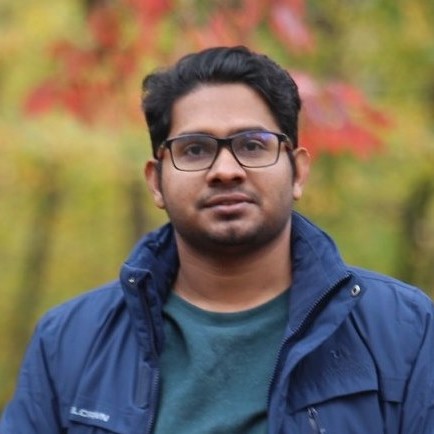}}]{M. Mahbubur Rahman} received the B.S. degree in Electronics and Communication Engineering from Khulna University of Engineering and Technology (KUET), Bangladesh, in 2016. He is currently a Ph.D. student in Electrical and Computer Engineering at the University of Alabama (UA), Tuscaloosa, AL, USA, and a research assistant in the UA Laboratory of Computational Intelligence for Radar (CI4R). His research interests include radar signal processing, machine learning, and multi-modal sensing for fall detection and gait analysis, vehicular autonomy and human-computer interaction.

M.M. Rahman is a recipient of the UA Graduate Council Fellowship in September 2019 and 3rd place in the Best Student Paper Competition of the IEEE Radar Conference in April 2021.
\end{IEEEbiography}

\begin{IEEEbiography}[{\includegraphics[width=1in,height=1.25in,clip,keepaspectratio]{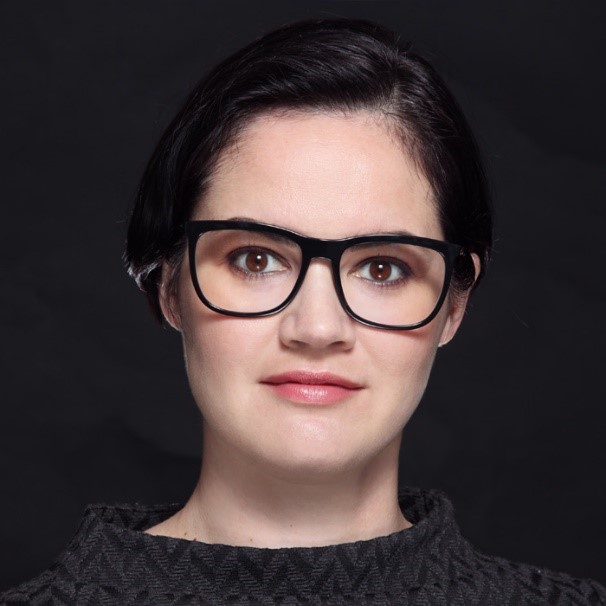}}]{Evie A. Malaia} received her Ph.D. degree in Computational Linguistics from Purdue University, West Lafayette, in 2004.  

Formerly a Research Scientist at Indiana University and Purdue University, and an Assistant Professor at the University of Texas at Arlington, she is currently an Associate Professor at the University of Alabama at Tuscaloosa, Department of Communicative Disorders.  Her current research interests include neural and physical bases of sign language communication, classification of higher cognitive states, and neural bases of autism spectrum disorders.

Dr. Malaia is a recipient of the Ralph E. Powe Award from DOE/ORAU, EurIAS Research Fellowship, EU Marie Curie Senior Research Fellowship, and the APS Award for Teaching and Public Understanding of Psychological Science.
\end{IEEEbiography}

\begin{IEEEbiography}[{\includegraphics[width=1in,height=1.25in,clip,keepaspectratio]{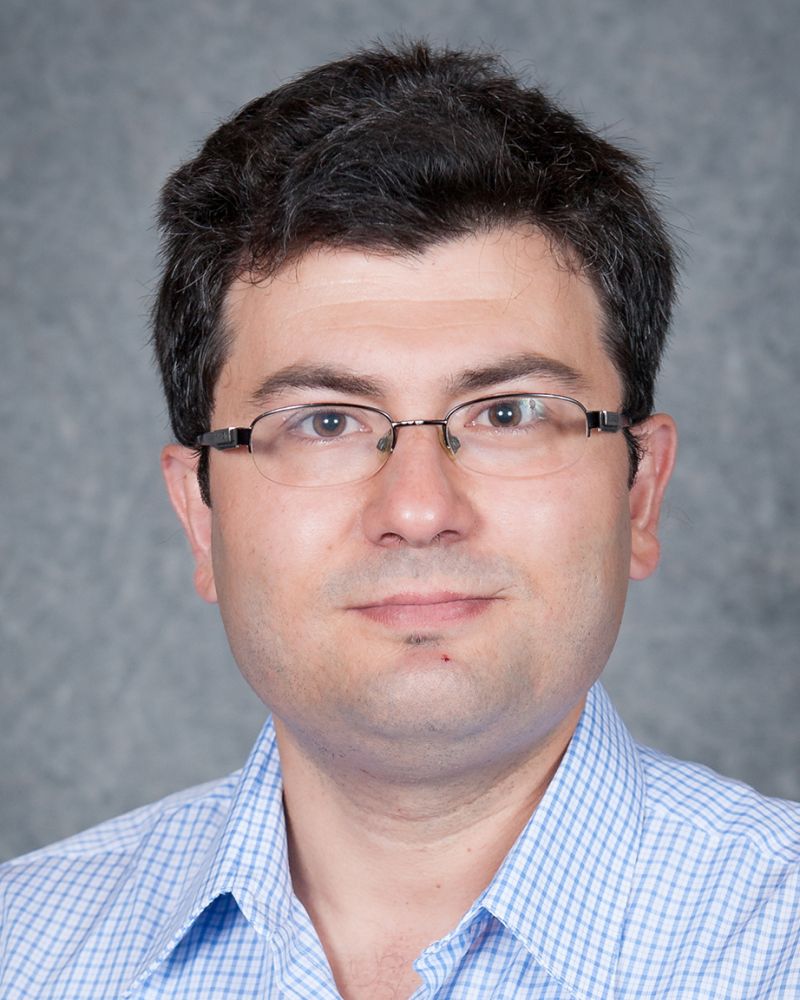}}]{Ali Cafer Gurbuz} received B.S. degree from Bilkent University, Ankara, Turkey, in 2003, in Electrical Engineering, and the M.S. and Ph.D. degrees from Georgia Institute of Technology, Atlanta, GA, USA, in 2005 and 2008, both in Electrical and Computer Engineering. From 2003 to 2009, he researched compressive sensing based computational imaging problems at Georgia Tech. He held faculty positions at TOBB University and University of Alabama between 2009 and 2017 where he pursued an active research program on the development of sparse signal representations, compressive sensing theory and applications, radar and sensor array signal processing, and machine learning.  Currently, he is an Assistant Professor at Mississippi State University, Department of Electrical and Computer Engineering, where he is co-director of Information Processing and Sensing (IMPRESS) Lab. 

Dr. Gurbuz is the recipient of The Best Paper Award for Signal Processing Journal in 2013 and the Turkish Academy of Sciences Best Young Scholar Award in Electrical Engineering in 2014. He has served as an associate editor for several journals such as Digital Signal Processing, EURASIP Journal on Advances in Signal Processing and Physical Communications. 
\end{IEEEbiography}

\begin{IEEEbiography}[{\includegraphics[width=1in,height=1.25in,clip,keepaspectratio]{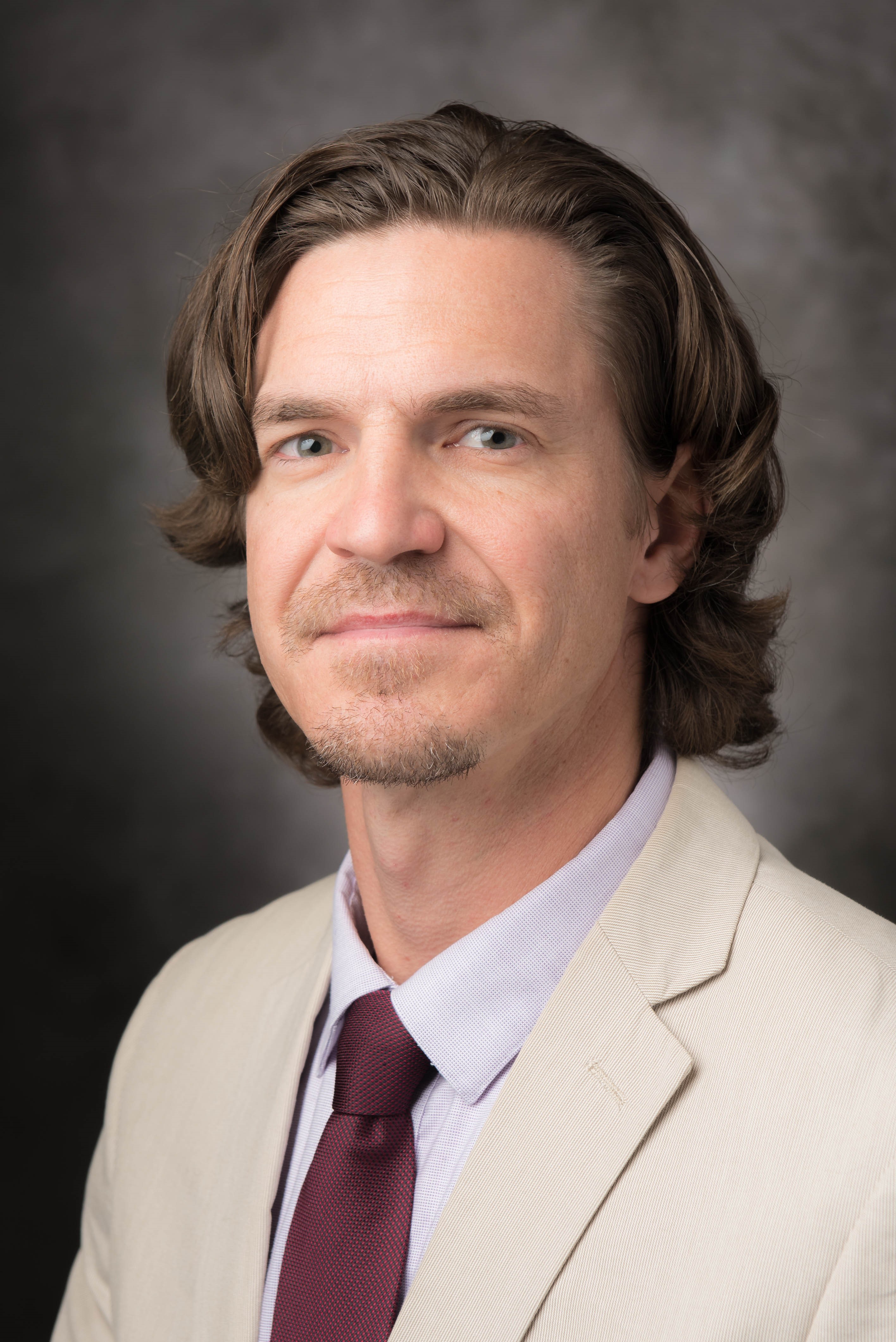}}]{Darrin J. Griffin} received the B.S. degree in communication sciences and disorders with a focus on deaf education and the M.A. degree in communication studies in 2004 and 2007, respectively, from The University of Texas at Austin. The Ph.D. degree was completed at The University at Buffalo, SUNY in 2010 in communication with a focus on deceptive communication.

From August 2010 to current he has served as a faculty member at The University of Alabama, Department of Communication Studies where he currently teaches and conducts research as an associate professor on topics related to nonverbal communication, deceptive communication, and deafness. Dr. Griffin is fluent in American Sign Language and participates in various forms of community engagement with the Deaf community.

Dr. Griffin is recipient of the 2020 College of Communication and Information Sciences Board of Visitors Research Excellence Award; the 2018 President’s Faculty Research Award at The University of Alabama; and a 2018 Premiere Award from The University of Alabama Council on Community-Based Partnerships for research that raised weather awareness and preparedness for the Deaf \& hard of hearing community. 

\end{IEEEbiography}

\begin{IEEEbiography}[{\includegraphics[width=1in,height=1.25in,clip,keepaspectratio]{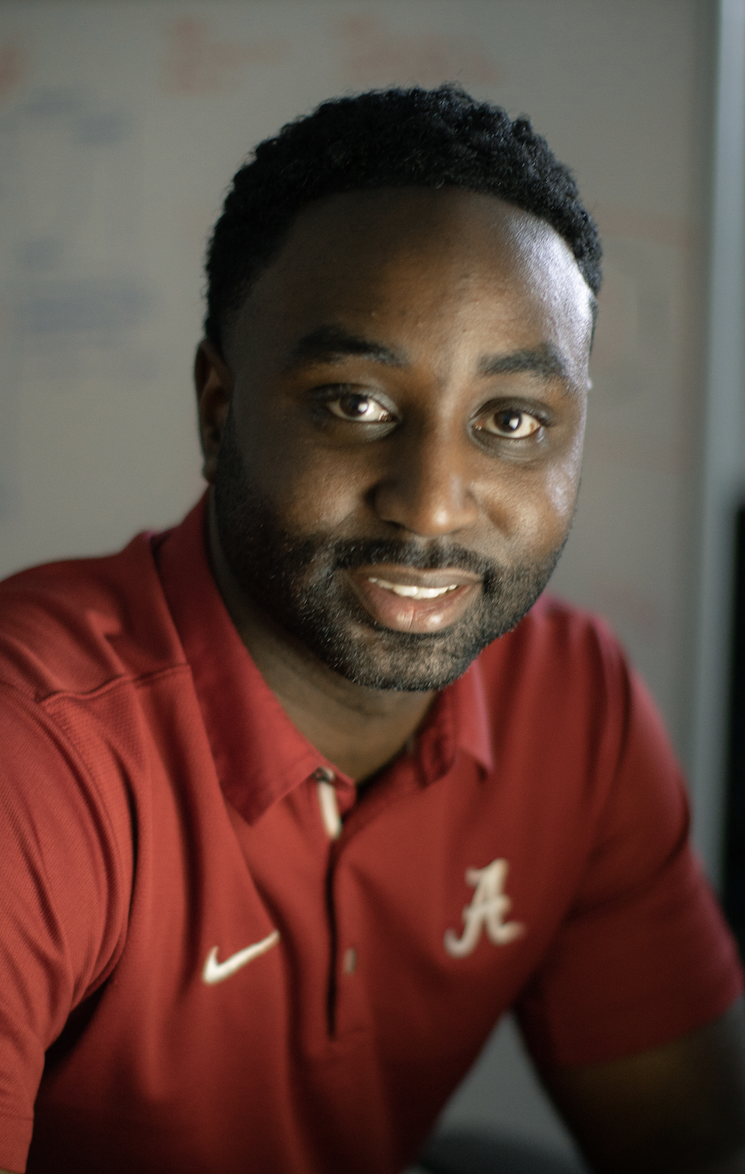}}]{Chris S. Crawford} received the Ph.D. degree in human-centered computing from the University of Florida, Gainesville, FL, USA. He is currently an Assistant Professor at the University of Alabama’s Department of Computer Science. He directs the Human-Technology Interaction Lab (HTIL). He has investigated multiple systems that provide computer applications and robots with information about a user’s cognitive state. In 2016, he lead the development of a BCI application that was featured in the world's first multiparty brain-drone racing event.  His current research focuses on computer science education, human-robot interaction, and brain-computer interfaces.
\end{IEEEbiography}

\begin{IEEEbiography}[{\includegraphics[width=1in,height=1.25in,clip,keepaspectratio]{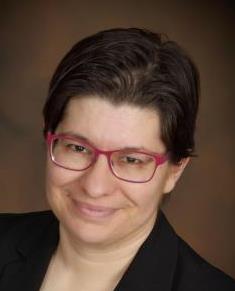}}]{Sevgi Z. Gurbuz} (S'01--M'10--SM'17) received the B.S. degree in electrical engineering with minor in mechanical engineering and the M.Eng. degree in electrical engineering and computer science from the Massachusetts Institute of Technology, Cambridge, MA, USA, in 1998 and 2000, respectively, and the Ph.D. degree in electrical and computer engineering from Georgia Institute of Technology, Atlanta, GA, USA, in 2009.  
     
From February 2000 to January 2004, she worked as a Radar Signal Processing Research Engineer with the U.S. Air Force Research Laboratory, Sensors Directorate, Rome, NY, USA.  Formerly an Assistant Professor in the Department of Electrical-Electronics Engineering at TOBB University, Ankara, Turkey and Senior Research Scientist with the TUBITAK Space Technologies Research Institute, Ankara, Turkey, she is currently an Assistant Professor in the University of Alabama at Tuscaloosa, Department of Electrical and Computer Engineering.  Her current research interests include physics-aware machine learning, RF sensor-enabled cyber-physical systems, radar signal processing, sensor networks, human motion recognition for biomedical, automotive autonomy, and human-computer interaction (HCI) applications.
     
Dr. Gurbuz is a recipient of the IEEE Harry Rowe Mimno Award for 2019, 2020 SPIE Rising Researcher Award, EU Marie Curie Research Fellowship, and the 2010 IEEE Radar Conference Best Student Paper Award.
\end{IEEEbiography}

\bibliographystyle{IEEEtran}
\bibliography{gurbuzrefs,refs,ASLR_PublicationBibtex}

\end{document}